\documentclass[aps,prl,floatfix,twocolumn,nofootinbib,superscriptaddress,groupedaddress]{revtex4-2}

\usepackage{amsmath,amssymb,bm,color,float,graphicx,mathrsfs}
\usepackage[hidelinks]{hyperref}
\hypersetup{colorlinks=true,linkcolor=blue,citecolor=blue,urlcolor=blue}
\usepackage{Macro}

\newcommand{\compile}{}
\newcommand{\submitletter}{}

\begin{document}

\ifdefined\compile

% Title %
\title{New probe of inflationary gravitational waves: cross-correlations of lensed primary CMB B-modes with large-scale structure}

\author{Toshiya Namikawa}
\affiliation{Center for Data-Driven Discovery, Kavli Institute for the Physics and Mathematics of the Universe (WPI), UTIAS, The University of Tokyo, Kashiwa, 277-8583, Japan}
\author{Blake D. Sherwin}
\affiliation{Department of Applied Mathematics and Theoretical Physics, University of Cambridge, Wilberforce Road, Cambridge CB3 0WA, United Kingdom}
\affiliation{Kavli Institute for Cosmology, University of Cambridge, Madingley Road, Cambridge CB3 OHA, United Kingdom}

% Date %
\date{\today}

% Abstract %
\begin{abstract}
We propose a new probe of inflationary gravitational waves (IGWs): the cross-correlation of the lensing of inflationary $B$-mode polarization with a large-scale structure (LSS) tracer, which can also be a cosmic microwave background (CMB) lensing map. This is equivalent to measuring a three-point function of two CMB $B$-modes and an LSS tracer. 
We forecast expected $1\,\sigma$ constraints on the tensor-to-scalar ratio $r$, albeit with a simplistic foreground treatment, and find constraints of $\sigma_r \simeq 7 \times 10^{-3}$ from the correlation of CMB-S4-Deep $B$-mode lensing and LSST galaxies, $\sigma_r \simeq 5 \times 10^{-3}$ from the correlation of CMB-S4-Deep $B$-mode lensing and CMB-S4-Deep CMB lensing, and $\sigma_r \simeq 10^{-2}$ from the correlation of LiteBIRD $B$-mode lensing and CMB-S4-Wide lensing. 
Because this probe is inherently non-Gaussian, simple Gaussian foregrounds will not produce any biases to the measurement of $r$. 
While a detailed investigation of non-Gaussian foreground contamination for different cross-correlations will be essential, 
this observable has the potential to be a useful probe of IGWs, which, due to different sensitivity to many potential sources of systematic errors, can be complementary to standard methods for constraining $r$.
\end{abstract} 

%////////////////////////////////////////%
% MAIN MATTER 
%////////////////////////////////////////%

% Contents %
\maketitle

\fi

%////////////////////////////////////////%
\ifdefined \submitletter
  {\it Introduction.---}
\else
  \section{Introduction} \label{sec:intro}
\fi
%////////////////////////////////////////%
% summary of current constraints on GW
Measurements of the cosmic microwave background (CMB) anisotropies have played a key role in developing the current picture in cosmology \cite{Komatsu:2014:wmap-review,Planck:2018:overview}. In the coming decades, measuring the polarization of the CMB will be at the forefront of observational cosmology. In particular, measurements of the parity-odd component -- the $B$-modes -- in the CMB polarization will be of great importance \cite{Kamionkowski:1996:GW,Seljak:1996:GW}, as these provide us with a unique avenue to test the presence of gravitational waves predicted by cosmic inflation \cite{Guth:1981,Sato:1981,Linde:1982,Albrecht:1982} and gain new insights into the early Universe (see reviews \cite{Kamionkowski:2015yta,Komatsu:2022:review} and references therein). 
Observations have not yet confirmed the presence of these inflationary gravitational waves (IGWs) but have placed upper bounds on the IGW amplitude, which is parametrized by the tensor-to-scalar ratio $r$, as $r\alt0.03$ ($2\,\sigma$)~\cite{BK13,Tristram:2021,Campeti:2022} at a pivot scale of $0.05\,$Mpc$^{-1}$, using data from BICEP/Keck Array, Planck, and WMAP. 
These measurements have already ruled out several inflationary models. 
Several ongoing and upcoming CMB experiments, including the BICEP Array \cite{BICEPArray}, Simons Array \cite{SimonsArray}, Simons Observatory (SO) \cite{SimonsObservatory}, LiteBIRD \cite{LiteBIRD:2022:PTEP}, and CMB-S4 \cite{CMBS4}, are targeting detections of, or much tighter constraints on, IGW $B$-modes over the next decade.

A high-precision measurement of the large-scale $B$-mode polarization has the potential to tightly constrain $r$ at the level of $\sigma_r\alt 10^{-3}$ in future CMB experiments \cite{CMBS4:r-forecast,LiteBIRD:2022:PTEP}. However, precise measurements of IGW $B$-modes must overcome bias and noise arising from other $B$-mode sources. One of the most challenging issues is Galactic foregrounds (e.g., \cite{B2I,BK-XVI:FG}). The large-scale $B$-modes are dominated by polarized Galactic foregrounds, so that we need to mitigate Galactic foregrounds very accurately. Multiple studies have presented increasingly powerful techniques for modeling or mitigating foregrounds (see, e.g., \cite{Eriksen:2007,Dunkley:2008,Betoule:2009,Katayama:2011eh,Ichiki:2014,Chluba:2017,Azzoni:2020,delaHoz:2020,Remazeilles:2020,Puglisi:2021,Delouis:2022,Minami&Ichiki:2022,Rahman:2022,Vacher:2022,Carones:2022}), but no observations have yet demonstrated bias-free foreground cleaning at the level of $r<10^{-2}$ in data. Foreground cleaning can also lead to a significant increase of statistical uncertainties in a $B$-mode measurement, especially if the foregrounds are very complex \cite{LiteBIRD:2022:PTEP}. 

Here, we propose a new way to search for IGWs, which has an entirely different -- and potentially lower -- sensitivity to Galactic foregrounds. 
The method utilizes the fact that the IGW $B$-modes are affected by the gravitational lensing effect induced by the large-scale structure (LSS), while the polarized foregrounds are not. 
More specifically, we propose to estimate a lensing signal that scales with $r$ from two observed $B$-modes, using the standard lensing quadratic estimator \citep{HuOkamoto:2001} but optimized for IGW $B$-modes.
Cross-correlations of this $r$-dependent lensing signal with the LSS then provide a constraint on $r$. 
This correlation is equivalent to the bispectrum between two $B$-modes and a LSS tracer. 
The correlation is immune to Galactic foregrounds if either (i) the measured LSS tracer does not contain significant imprints of Galactic foregrounds or (ii) the Galactic foregrounds can be well approximated as Gaussian since the bispectrum observable is fundamentally non-Gaussian. The LSS tracer can correspond to a CMB lensing map reconstructed from small-scale data or to a galaxy number density map, both of which may only have minimal contamination by Galactic foregrounds.
While the foregrounds are, of course, known to be non-Gaussian, we note that any remaining non-Gaussian foreground bias can be mitigated with a suitable choice of LSS tracers or CMB lensing maps as we will outline in {\it Discussion}. 

%////////////////////////////////////////%
\ifdefined \submitletter
  {\it Method.---}
\else
  \section{Method}
\fi
%////////////////////////////////////////%
We first begin with a brief reminder of the properties of CMB polarization anisotropies. CMB observations measure the Stokes $Q$ and $U$ parameters in each line-of-sight direction $\hatn$ on the unit sphere. We then define the scalar and pseudo-scalar $E$- and $B$-modes of the CMB polarization from the Stokes parameters as \cite{Zaldarriaga:1996:EBdef}
%--------------------------------------------------%
\al{
    E_{\l m} \pm\iu B_{\l m} = - \Int{2}{\hatn}{} (Y_{\l m}^{\pm 2}(\hatn))^* [Q\pm\iu U](\hatn)
    \,, 
}
%--------------------------------------------------%
where $Y_{\l m}^2(\hatn)$ are spin-2 spherical harmonics. 

The path of CMB photons is affected by the gravitational potential of the LSS along the geodesic. The lensing distorts the polarization map by remapping it with a deflection angle $\bm{d}(\hatn)$ (see \cite{Lewis:2006fu} for a review). This lensing distortion mixes the $E$- and $B$-modes at different angular scales. Introducing the lensing potential $\phi$ so that $\bm{d}=\bn\phi$, the $B$-modes are modified as follows \cite{Zaldarriaga:1998:LensB}:
%--------------------------------------------------%
\al{
    \tilde{B}_{\l m} &= B_{\l m}+\sum_{LM\l'm'}(-1)^m\Wjm{\l}{L}{\l'}{-m}{M}{m'} 
    \notag \\
    &\qquad \times \grad_{LM}(-\iu E_{\l'm'}W^{-}_{\l L\l'}+B_{\l'm'}W^{+}_{\l L\l'}) 
    \,, \label{Eq:B-x}
}
%--------------------------------------------------%
where $\phi_{LM}$ is the spherical harmonic coefficients of the lensing potential, and we ignore higher-order terms in $\grad$. We define
%----------------------------------------------------------------------%
\al{
    W^{\pm}_{\l L\l'} &\equiv - q^{\pm}_{\l L\l'} \sqrt{\frac{L(L+1)(2\l+1)(2L+1)(2\l'+1)}{4\pi}}   
    \notag \\
	&\qquad\times \frac{1}{2}\bigg[ \sqrt{(\l'-2)(\l'+3)} \Wjm{\l}{L}{\l'}{2}{1}{-3} 
    \notag \\
	&\qquad\qquad + \sqrt{(\l'+2)(\l'-1)} \Wjm{\l}{L}{\l'}{-2}{1}{1} 
	\bigg]
	\,,
}
%----------------------------------------------------------------------%
with $q^\pm_{\l L\l'}\equiv [1\pm (-1)^{\l+L+\l'}]/2$. 

We now propose our new approach to detect the IGWs. 
We first note that lensing breaks the statistical isotropy of the $B$-modes, leading to correlations between the lensed primary CMB $B$-modes at different angular scales:
%--------------------------------------------------%
\al{
    \ave{\tilde{B}_{\l m}\tilde{B}_{\l'm'}} = \sum_{LM}\Wjm{\l}{\l'}{L}{m}{m'}{M} f^\psi_{\l L\l'} r\phi^*_{LM}
    \,,\label{Eq:mode-coupling:BB}
}
%--------------------------------------------------%
where the ensemble average is taken over CMB realizations with a fixed $\phi$, and we introduce a response function
$f^\psi_{\l L\l'} = W^{+}_{\l L\l'}C^{BB,r=1}_{\l'} + (-1)^{\l+L+\l'}W^{+}_{\l'L\l}C^{BB,r=1}_\l$
(see also Ref.~\cite{Fabbian:2019} for the non-perturbative response in the flat-sky approximation). 
Here, $C_\l^{BB,r=1}$ is the IGW $B$-mode power spectrum with $r=1$. 
We therefore can use this correlation to reconstruct $\psi\equiv r\phi$. 
The reconstructed $\psi$ fields from the large-scale lensed $B$-modes alone are then cross-correlated with an external LSS tracer, $x$. 
This is equivalent to measuring a cross-bispectrum between two CMB $B$-modes and the LSS tracer, $x$. 
More specifically, for the idealistic full-sky case, we follow closely the construction of quadratic CMB lensing estimators in Ref.~\cite{OkamotoHu:quad} and define an estimator to reconstruct $\psi$ from the observed $B$-modes, $\hat{B}_{\l m}$, as
%--------------------------------------------------%
\al{
    \hat{\psi}^*_{LM}
    = \frac{1}{2}A_L\sum_{\l\l'mm'}\Wjm{\l}{\l'}{L}{m}{m'}{M}
    f^\psi_{\l L\l'}\frac{\hat{B}_{\l m}}{\hC^{BB}_\l}\frac{\hat{B}_{\l'm'}}{\hC^{BB}_{\l'}}
    \,, \label{Eq:estg} 
}
%--------------------------------------------------%
where $\hC^{BB}_\l$ is 
the best estimate of the observed power spectrum from theoretical computation or simulation. 
$A_L$ is the estimator normalization and is given in the idealistic full-sky case by
%--------------------------------------------------%
\al{
	A_L^{-1} = \frac{1}{2L+1}\sum_{\l\l'}\frac{(f^\psi_{\l L\l'})^2}{2\hC^{BB}_\l\hC^{BB}_{\l'}} 
	\,. \label{Eq:Norm:phi}
}
%--------------------------------------------------%
As defined above, $\ave{\hat{\psi}}=r\phi$ is proportional to $r$. 
The cross-power spectrum between $\hat{\psi}$ and an LSS tracer is therefore also proportional to $r$. Thus, we can constrain $r$ by measuring the amplitude of the cross-power spectrum $C_L^{\psi x}=rC_L^{\phi x}$ and, provided that $C_L^{\phi x}$ is well determined by other measurements, finding the best fit value of $r$. 
%The cross-power spectrum between $\hat{\psi}$ and a LSS tracer is therefore also proportional to $r$. Thus, we can constrain $r$ by measuring the amplitude of the cross-power spectrum, $C_L^{\psi x}=r C_L^{\phi x}$ and, assuming that the standard lensing spectrum $C_L^{\phi x}$ is well known from other measurements\footnote{This assumption is reasonable given that precise lensing maps and hence lensing cross- and auto-spectra will be available from upcoming experiments.}, finding the best fit value of $r$. 
%This approach is similar to the one used in \cite{Namikawa:2021:mode} to constrain cosmic birefringence using the odd-parity bispectrum. 

The observed $B$-modes in Eq.~\eqref{Eq:estg} suffer from a large cosmic variance due to the presence of $B$-modes converted from $E$-modes by lensing, i.e., the first term in the second line of Eq.~\eqref{Eq:B-x}. To reduce this variance, we consider applying the following template-delensing method. One can create a template of the lensing-induced $E$-to-$B$ leakage using an estimate of lensing potential $\hat{\phi}$ and observed $E$-modes, and then subtract this template from the observed $B$-modes (e.g., \cite{Seljak:2003pn,BaleatoLizancos:2020:delens-bias}) as follows\footnote{Note that $\hat{\phi}$ can either derive from CMB lensing reconstruction or from a suitably scaled LSS tracer as in \cite{Sherwin:2015}.}:
\al{
    \hat{B}^{\rm del}_{\l m} &= \hat{B}_{\l m}-\sum_{LM\l'm'}(-1)^m\Wjm{\l}{L}{\l'}{-m}{M}{m'} 
    \notag \\
    &\times \left(\frac{C^{\phi\phi}_{L} \hat{\grad}_{LM}} {C^{\phi\phi}_{L} + N^{\phi\phi}_{L} }\right)\left(\frac{-\iu C^{EE}_{\l'} \hat{E}_{\l'm'}}{C^{EE}_{\l'} + N^{EE}_{\l'} }\right)W^{-}_{\l L\l'} 
    \,. \label{Eq:B-del}
}
Here, $N^{\phi\phi}_L$ and $N^{EE}_\l$ are the noise power spectra of the lensing estimate and $E$-modes, respectively. 
The delensed modes $\hat{B}^{\rm del}_{\l m}$ and their power spectra are then used for the $\psi$-reconstruction with the estimator of Eq.~\eqref{Eq:estg}. To leading order in $\grad$, this 
delensing method only reverses lensing of $E$-modes but, crucially, does not undo the lensing of IGW $B$-modes in Eq.~\eqref{Eq:B-x}; 
%\footnote{In contrast, delensing by inverse-remapping the CMB polarization (e.g., \cite{Carron:2017,Carron:2018}) will not work well, as it will also undo the lensing of the IGW $B$-modes, which is the focus of our method.} 
it hence does not affect the mode coupling of Eq.~\eqref{Eq:mode-coupling:BB} that allows us to reconstruct $\psi$. To a good approximation, the method just reduces the residual lensing $B$-mode power that contributes to $\hC^{BB}_\l$ by a factor $A_{\rm lens}$ and thus lowers the reconstruction noise of the estimator of Eq.~\eqref{Eq:estg}. 

The expected $1\,\sigma$ uncertainty of $r$ from the cross-correlations between $\psi$ and an LSS tracer is
\al{
    \sigma^{-2}_r 
    &= \sum_L \frac{(2L+1)f_{\rm sky}(\pd C_L^{\psi x}/\pd r)^2}{(r^2C_L^{\phi\phi}+N_L^{\psi\psi})\hC_L^{xx}+(rC_L^{\phi x})^2}
    \,, 
}
where $f_{\rm sky}$ is the sky fraction of a CMB polarization observation, $\hC_L^{xx}$ is the observed power spectrum of $x$, and $N_L^{\psi\psi}$ is the reconstruction noise spectrum. 
For small fiducial values of $r$, 
%(which for the forecast configurations considered in our analysis correspond to values of $r\ll 0.1$), 
the above equation can be approximated as 
\al{
    \sigma^{-2}_r \simeq \sum_L (2L+1)f_{\rm sky}\frac{\rho^2_LC_L^{\phi\phi}}{N_L^{\psi\psi}}
    \,, \label{Eq:sigmar}
}
where $\rho_L\equiv C_L^{\phi x}/\sqrt{C^{\phi\phi}_L\hC^{xx}_L}$ is the correlation coefficient between the true CMB lensing potential and the observed LSS tracer. 
In the idealized fullsky case, the reconstruction noise is equal to the estimator normalization and we adopt $N_L^{\psi\psi}\simeq A_L$. 
Note that the cosmic variance of the IGW $B$-modes increases $N_L^{\psi\psi}$ as in Eq.~\eqref{Eq:Norm:phi}. In our analysis, we assume $r=0$ and focus on the significance of rejecting the null hypothesis.

%////////////////////////////////////////%
\ifdefined \submitletter
  {\it Forecast.---}
\else
  \section{Forecast} \label{sec:results}
\fi
%////////////////////////////////////////%
We first estimate $\sigma_r$ for specific experimental specifications. We assume that the different experiments listed in Table \ref{tab:experiments:psi} (SO-SAT, LiteBIRD, S4D-SAT) measure $\psi$. Table \ref{tab:experiments} summarizes our results, where we assume that the experiments in the second column of Table \ref{tab:experiments} (SO-LAT, S4W, S4D-LAT, LSST galaxies) provide the LSS tracer $x$ used for cross-correlation with $\psi$. Note that ``SO," ``LiteBIRD," ``S4W," and ``S4D" mimic the Simons Observatory, LiteBIRD, CMB-S4 Wide Survey, and CMB-S4 Ultra-Deep Survey, respectively. To compute $\sigma_r$, we evaluate $N_L^{\psi\psi}$ and $\rho_L$ in Eq.~\eqref{Eq:sigmar} as follows.

To compute $N_L^{\psi\psi}$, we use a CMB noise spectrum with a polarization noise level $\sigma_{\rm P}$ and include the effect of beam deconvolution with angular resolution $\theta$.
We include in our forecasts the degradation of the noise level due to component separation as in Ref.~\cite{Errard:2016:FG} for LiteBIRD and S4. For SO-SAT, we use the noise curve corresponding to the goal noise level for the optimistic $1/f$ case shown in Fig.~11 of Ref.~\cite{SimonsObservatory}. We set the lowest multipole of the S4D-SAT to $50$, corresponding to the $\ell_{\rm knee}$ of the $1/f$ noise \cite{CMBS4:r-forecast}.
We note that the residual foregrounds increase the variance of the large-scale $B$-modes and affect the $\psi$-reconstruction noise. In component separation with suitable algorithms, however, the residual foregrounds can typically be suppressed to be small compared to the noise (e.g., \cite{Errard:2016:FG}), and so we neglect these residuals in our forecasts except for LiteBIRD where we include the residual foregrounds of Ref.~\cite{Errard:2016:FG}. 
We also include delensing prior to the $\psi$ measurement by simply scaling the overall amplitude of the lensing $B$-mode power by a value $A_{\rm lens}$, which describes the appropriate level of residual $B$-modes after delensing each experiment.

To compute $\rho_L$, we consider either reconstructed $\phi$ or galaxy number density. For the $\phi$ measurement, since the large-scale $B$-modes from a high-resolution experiment usually suffer from high atmospheric $1/f$ noise levels, we do not use $B$-modes at $\l<500$ for computing the $\phi$-reconstruction noise. 
This split also avoids any delensing bias arising from the correlation between the $B$-modes in the $\phi$-reconstruction and those to be delensed (e.g., \cite{Teng:2011xc,Namikawa:2014:patchwork,BaleatoLizancos:2020:delens-bias}).
For $E$-modes used for the $\phi$-reconstruction, we use $\l\geq 100$ where the CMB signal is dominant \cite{SimonsObservatory}. 
We compute the $\phi$-reconstruction noise using the iterative approach to mimic optimal lensing reconstruction \cite{Smith:2010gu}. 

\begin{table}[t]
\centering
\begin{tabular}{l|ccccc}
    Experiment for $\psi$ & $\sigma_{\rm P}$ ($\mu$K$'$) & $\theta$ ($'$) & $\l_{\rm min}$ & $\l_{\rm max}$ & $A_{\rm lens}$ \\ \hline
    SO-SAT &  &  & 30 & 300 & 0.3 \\
    LiteBIRD & 2.6 & 30 & 2 & 500 & 0.2 \\
    S4D-SAT & 0.5 & 20 & 50 & 500 & 0.05 \\
\end{tabular}
\caption{Specific experimental setup for a $\psi$ measurement used in our forecasts. For each experiment, the noise level includes degradation due to the component separation and the remaining fraction of the lensing $B$-mode power parametrized with $A_{\rm lens}$. For SO-SAT, we use the noise curve corresponding to the goal noise level for the optimistic $1/f$ case shown in Fig.11 of Ref.~\cite{SimonsObservatory}. $A_{\rm lens}$ for SO-SAT, LiteBIRD, and S4D-SAT are derived by assuming delensing of Ref.~\cite{Namikawa:2021:so-delens}, S4W, and S4D internal, respectively.}
\label{tab:experiments:psi}
\end{table}

\begin{table}[t]
\centering
\begin{tabular}{cc|c|c}
    $\psi$ & $x$ & $f_{\rm sky}$ & $\sigma_r$ \\ \hline
    SO-SAT & SO-LAT $\phi$ & 0.1 & $0.04$ \\ 
    LiteBIRD & S4W $\phi$ & 0.4 & $0.01$ \\ 
    S4D-SAT & S4D-LAT $\phi$ & 0.03 & $0.005$ \\
    S4D-SAT & LSST galaxies & 0.03 & $0.007$
\end{tabular}
\caption{Forecasts for $1\sigma$ constraints on $r$ using $C_L^{\psi x}$ at $L$ in $[10,500]$. We choose $f_{\rm sky}$ to only include the overlap regions between experiments. We choose $\sigma_{\rm P}=6\,\mu$K$'$ for SO-LAT, $1\,\mu$K$'$ for S4W, and $0.4\,\mu$K$'$ for S4D-LAT. We assume $\theta=1'$ for all these CMB experiments and use multipoles between $\ell=100$ and $4000$ but remove $\ell<500$ for $B$-modes used for reconstructing $\phi$ to be correlated with $\psi$. For the LSST galaxies, we assume a constant correlation coefficient with lensing of $70\%$.}
\label{tab:experiments}
\end{table}

We find that, for S4D, $\sigma_r\simeq 5\times10^{-3}$. The constraints will be somewhat degraded if we use, as an LSS tracer, galaxies that have a lower correlation coefficient with the CMB lensing map instead of the lensing reconstruction map. However, this may be worthwhile because such a cross-correlation might reduce potential concerns from foregrounds.
For LSST galaxies \cite{LSST}, the correlation coefficient is expected to be $\rho_L\agt 0.7$ at $L\alt1000$ (e.g., \cite{Namikawa:2021:so-delens}); then the constraint becomes $\sigma_r\simeq 7\times10^{-3}$. 

Note that the high sensitivity of our method to $r$ comes first from the fact that the estimator is not limited by the cosmic variance until we detect nonzero IGW $B$-modes, and second, from the fact that cross-correlations can boost the signal-to-noise ratio if only one of the fields being correlated has high noise.

Since the sensitivity to $r$ might be further improved by optimizing the experimental setup, including the number of frequency bands and detectors currently assumed in each experiment, we next discuss how $\sigma_r$ changes if we vary instrumental specifications. 
The constraint depends on at least four factors: (1) the observed sky fraction, (2) the correlation coefficient of the measured LSS tracer to $\phi$, (3) the noise level of the large-scale $B$-modes after component separation, and (4) the delensing efficiency. The dependence on $f_{\rm sky}$ is trivial. The correlation coefficients become close to unity if we use a CMB-S4 lensing map and are not much lower if we use a future, high-$z$ LSS survey. The last two determine the reconstruction noise of $\psi$, which enters into the denominator of Eq.~\eqref{Eq:sigmar}. 

\begin{figure}[t]
 \centering
 \includegraphics[width=86mm]{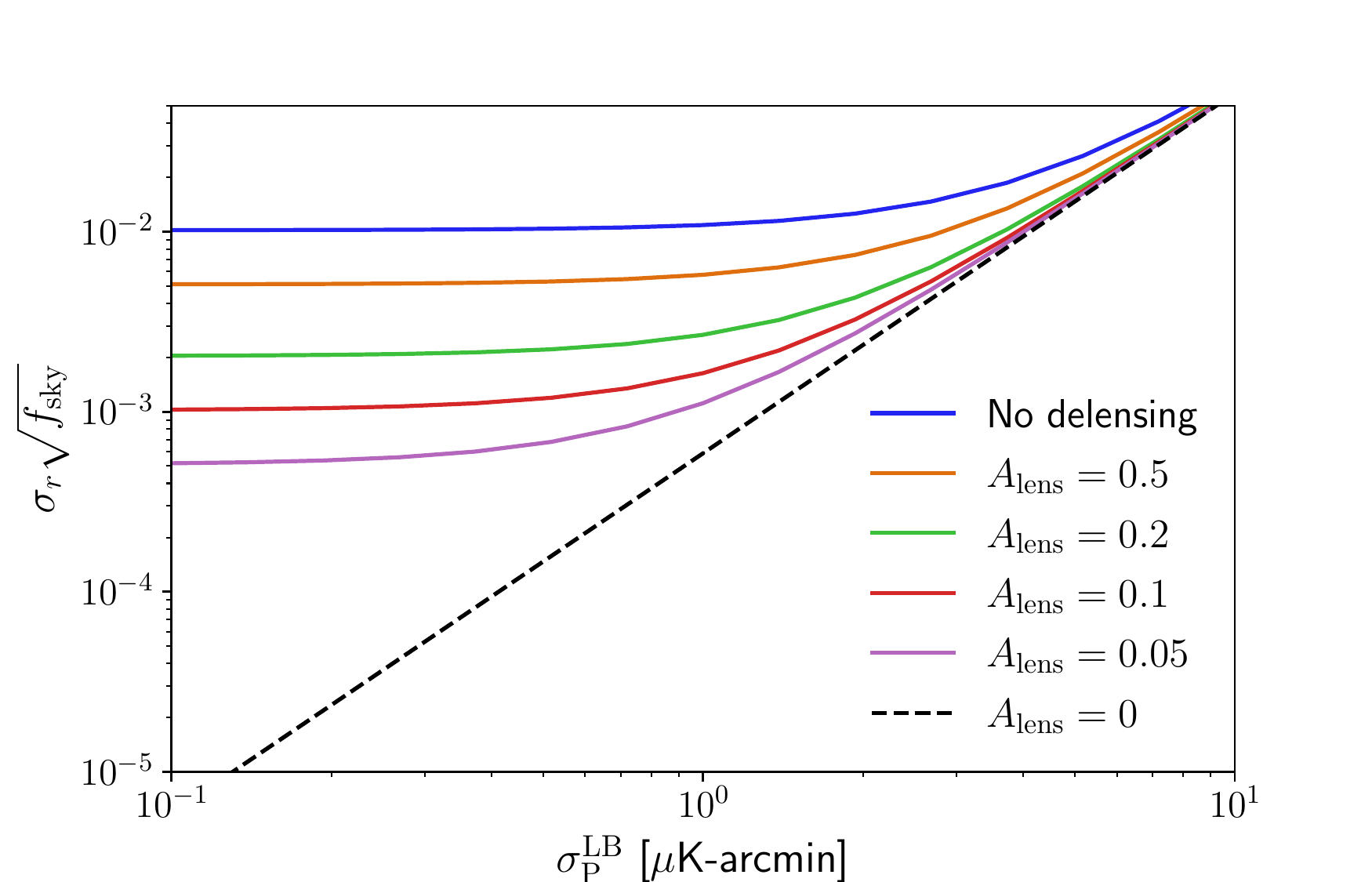}
 \caption{The $1\sigma$ constraints on $r$ arising from our method; we consider an experimental setup that resembles LiteBIRD+S4W but allows for a variation of the polarization noise level of LiteBIRD, $\sigma_{\mathrm{P}}^{\mathrm{LB}}$ and the delensing efficiency, parametrized via $A_{\rm lens}$. We use a reconstructed lensing map from S4W as an LSS tracer for the cross-correlation, although the results are not very sensitive to the detailed properties of this tracer. Note that $A_{\rm lens}=0.05$ corresponds to the S4D case, although $f_{\rm sky}$ and the noise level are allowed to vary here. We conclude that our method is capable of producing competitive constraints on $r$, especially if efficient delensing can be implemented.}
 \label{fig:const-r:sigS}
\end{figure}

Figure \ref{fig:const-r:sigS} shows $\sigma_r$ for a configuration that resembles the LiteBIRD+S4W setup except that we now allow the post-component-separation noise level of LiteBIRD $\sigma_{\rm P}^{\rm LB}$ and the delensing efficiency to vary. These two parameters change the performance of the $\psi$-reconstruction. Since the variance from the residual foregrounds is much smaller than that from the noise, we ignore the residual foregrounds in the calculation for simplicity. We assume the S4W $\phi$ map as an LSS tracer for cross-correlation, but the improvements when using a perfect $\phi$ map are negligible. On the other hand, a precise $\phi$ map is crucial for delensing: The constraints cannot improve without delensing as the noise level decreases below $\sigma_{\rm P}^{\rm LB}\ll 1\mu$K$'$ due to the lensing $B$-mode cosmic variance. 

We also explore the dependence of $\sigma_r$ on the multipole range of the large-scale $B$-modes used in the $\psi$-reconstruction. We find that removing $B$-modes below $\l\alt 30$ or above $\l\agt 200$ only increases $\sigma_r$ by $10\%$ for the LiteBIRD+S4W case. Even removing multipoles below $\l\alt 60$ or above $\l\agt 100$ only increases $\sigma_r$ by $50\%$. For S4D configurations, our findings are similar: Assuming the standard minimum multipole for the S4D setup, $\l_{\rm min}=50$, only increases $\sigma_r$ by $\sim 20\%$ compared to the case when multipoles $\l<50$ are included. 
This is because the lensing becomes important at higher multipoles but then the noise (and lensing noise) become dominant, so the recombination bump ($60<\l<100$) contains the most information on the lensing of the IGW $B$-modes.
Our method thus does not require $B$-mode information from the largest (or smallest) angular scales that can be most challenging for CMB experiments to observe and foreground-clean. 

%////////////////////////////////////////%
\ifdefined \submitletter
  {\it Discussion.---}
\else
  \section{Discussion} \label{sec:discussion}
\fi
%////////////////////////////////////////%
% Comparison with BB
We first discuss comparison with $r$ constraints from the standard analysis using the $B$-mode autopower spectrum. For example, if we assume the LiteBIRD+S4W case with $A_{\rm lens}=0.2$, the statistical uncertainty from the standard method in the same analysis setup is $\sigma_r\simeq2\times 10^{-4}$, which is more than an order of magnitude better than our method. However, the standard method suffers from a large bias from residual Galactic foregrounds (e.g., \cite{Aurlien:2022:FG}). A more realistic estimate by \cite{LiteBIRD:2022:PTEP} reports $\sigma_r\simeq 10^{-3}$ with a conservative method to reduce the level of residual foreground biases and instrumental systematics. Since Gaussian foregrounds do not bias our estimator, this level of conservatism may not be required for our method, so that our simple forecasts may be more realistic. Moreover, we note that in the standard method the residual foreground bias can depend sensitively on the assumptions in the foreground model, so that a small mismatch between assumptions and real data can lead to a significant bias in the $r$ estimate. Thus, our new method will provide a valuable, independent test of constraints on $r$.

% non-Gaussian FGs
Our estimator could be biased by the foreground non-Gaussianity. However, tracers can be chosen to minimize the non-Gaussian foregrounds. For example, we can select galaxy catalogs to minimize the impact of extinction, or use a $\phi$ map measured from small-scale CMB fluctuations with subdominant or near-independent foreground fluctuations that are negligibly correlated with large-scale foregrounds. Furthermore, different LSS tracers can be used to cross-check foreground stability and gain confidence in the robustness of the results. Although much further work on this subject is needed before conclusions can be drawn, it seems plausible that foreground contamination could be much less limiting for our method than for the standard method. Even if foreground biases are not immediately negligible, multifrequency foreground cleaning could certainly also be used. In any case, our method will provide a complementary cross-check of standard results, with any sensitivity to Galactic foregrounds entering in a qualitatively different manner.

Polarized extragalactic foregrounds -- in particular, polarized radio sources -- could be an additional concern; if the sources contributed a $B$-mode signal $S_B$, they could, in principle, bias our measurement by a bispectrum contribution $S_B S_B \phi$. The delensing process also potentially introduces a bias \cite{BaleatoLizancos:2022:extFGs}. Fortunately, we can use standard techniques from lensing reconstruction to mitigate potential point source biases. In particular, we can combine aggressive source masking with modifications of the estimator, using analogs of bias-hardening in lensing reconstruction \cite{Namikawa:2013:bhepol,Sailer:2022}, to further mitigate any foreground biases.

% non-Gaussian covariance
We note that our analysis did not include the non-Gaussian covariance of the bispectrum. The $B$-modes are non-Gaussian due to the lensing $B$-modes even after foreground cleaning. However, the large-scale lensing $B$-modes are well described by a Gaussian field \cite{Namikawa:2015:delens}, and such non-Gaussianity should hence be small. 

% higher-order bias from phi
We have also ignored terms arising from higher orders in $\phi$. While we defer a detailed analysis to future work, we will discuss this issue briefly and schematically here. By expanding the delensed $B$-modes to higher orders in $\phi$ and substituting this expansion into our $\psi$ estimator, we find that the leading higher-order terms in the cross-correlation are either: i) terms such as (schematically) $\sim C_\l^{EE}A^{1/2}_{\rm lens}\phi^4$ and $\sim C_\l^{EE}A_{\rm lens}\phi^3$ that are proportional to $A_{\rm lens}$ and hence should be substantially reduced by efficient delensing, or ii) terms such as $\sim N_\l^{EE}\phi^3$ that arise from $E$-mode noise and can be removed by cross-correlation of different splits of the data. Although these terms may not be entirely negligible, we therefore expect them to be small enough that any bias they produce could simply be modeled and subtracted.

%////////////////////////////////////////%
\ifdefined \submitletter
  {\it Summary.---}
\else
  \section{Summary} \label{sec:summary}
\fi
%////////////////////////////////////////%
We have proposed a new method to measure $r$ using cross-correlations of an LSS tracer with lensing of primary $B$-modes, reconstructed with a quadratic estimator. We have explored the sensitivity of this cross-correlation to IGWs for upcoming CMB experiments and have found that the cross-correlation can constrain $r$ with $\sigma_r\simeq 5\times 10^{-3}-4\times 10^{-2}$. Compared to the $B$-mode power spectrum, our method is insensitive to Gaussian foregrounds and would be less limited by Galactic foreground uncertainties more generally. Our method can thus potentially constrain $r$ in a manner that is complementary to standard analyses with the $B$-mode autopower spectrum.

In closing, we note a number of interesting features of our method. 
%Since the covariance between the standard $B$-mode power spectrum and our cross-correlation is a five-point correlation and should thus be negligible, $\sigma_r$ can be further improved by combining the two statistics. 
Compared to standard $B$-mode power spectrum analyses, our method is expected to have somewhat different sensitivity to the spectral shape of the IGW power spectrum. Furthermore, in principle, our method can also probe a redshift dependence of $B$-mode sources by using a tomographic measurement of the cross-correlation with LSS tracers at different redshifts. We will discuss these points in more detail in our follow-up papers.
%Since the covariance between the standard $B$-mode power spectrum and our cross-correlation is a five-point correlation and should thus be negligible, $\sigma_r$ can be further improved by combining the two statistics. Furthermore, compared to standard $B$-mode power spectrum analyses, our method is expected to have somewhat different sensitivity to the spectral shape of the IGW power spectrum. Finally, in principle, our method can also probe a redshift dependence of $B$-mode sources by using a tomographic measurement of the cross-correlation with LSS tracers at different redshifts. We will discuss these points in more detail in our follow-up papers.

%////////////////////////////////////////%
% BACK MATTER 
%////////////////////////////////////////%

\begin{acknowledgments}
We thank Anton Baleato-Lizancos, Anthony Challinor, William R. Coulton, Colin Hill, Eiichiro Komatsu, Antony Lewis, Mathew Madavacheril, and Tomotake Matsumura for helpful comments and discussions. 
This work is supported in part by JSPS KAKENHI Grants No. JP20H05859 and No. JP22K03682 (T.N.). Part of this work uses the resources of the National Energy Research Scientific Computing Center. The Kavli IPMU is supported by World Premier International Research Center Initiative (WPI Initiative), MEXT, Japan.
\end{acknowledgments}

\appendix

% supplementary

% References %
\bibliographystyle{apsrev}
\bibliography{cite}

\begin{thebibliography}{56}
\expandafter\ifx\csname natexlab\endcsname\relax\def\natexlab#1{#1}\fi
\expandafter\ifx\csname bibnamefont\endcsname\relax
  \def\bibnamefont#1{#1}\fi
\expandafter\ifx\csname bibfnamefont\endcsname\relax
  \def\bibfnamefont#1{#1}\fi
\expandafter\ifx\csname citenamefont\endcsname\relax
  \def\citenamefont#1{#1}\fi
\expandafter\ifx\csname url\endcsname\relax
  \def\url#1{\texttt{#1}}\fi
\expandafter\ifx\csname urlprefix\endcsname\relax\def\urlprefix{URL }\fi
\providecommand{\bibinfo}[2]{#2}
\providecommand{\eprint}[2][]{\url{#2}}

\bibitem[{\citenamefont{Komatsu and Bennett}(2014)}]{Komatsu:2014:wmap-review}
\bibinfo{author}{\bibfnamefont{E.}~\bibnamefont{Komatsu}} \bibnamefont{and}
  \bibinfo{author}{\bibfnamefont{C.~L.} \bibnamefont{Bennett}}
  (\bibinfo{collaboration}{WMAP Science Team}), \bibinfo{journal}{\ptep}
  \textbf{\bibinfo{volume}{2014}}, \bibinfo{pages}{06B102}
  (\bibinfo{year}{2014}), \eprint{1404.5415}.

\bibitem[{\citenamefont{{\textit{Planck}
  Collaboration}}(2020)}]{Planck:2018:overview}
\bibinfo{author}{\bibnamefont{{\textit{Planck} Collaboration}}},
  \bibinfo{journal}{\aap} \textbf{\bibinfo{volume}{641}}, \bibinfo{pages}{A1}
  (\bibinfo{year}{2020}), \eprint{1807.06205}.

\bibitem[{\citenamefont{Kamionkowski et~al.}(1997)\citenamefont{Kamionkowski,
  Kosowsky, and Stebbins}}]{Kamionkowski:1996:GW}
\bibinfo{author}{\bibfnamefont{M.}~\bibnamefont{Kamionkowski}},
  \bibinfo{author}{\bibfnamefont{A.}~\bibnamefont{Kosowsky}}, \bibnamefont{and}
  \bibinfo{author}{\bibfnamefont{A.}~\bibnamefont{Stebbins}},
  \bibinfo{journal}{\prl} \textbf{\bibinfo{volume}{78}}, \bibinfo{pages}{2058}
  (\bibinfo{year}{1997}), \eprint{astro-ph/9609132}.

\bibitem[{\citenamefont{Seljak and Zaldarriaga}(1997)}]{Seljak:1996:GW}
\bibinfo{author}{\bibfnamefont{U.}~\bibnamefont{Seljak}} \bibnamefont{and}
  \bibinfo{author}{\bibfnamefont{M.}~\bibnamefont{Zaldarriaga}},
  \bibinfo{journal}{\prl} \textbf{\bibinfo{volume}{78}}, \bibinfo{pages}{2054}
  (\bibinfo{year}{1997}), \eprint{astro-ph/9609169}.

\bibitem[{\citenamefont{Guth}(1981)}]{Guth:1981}
\bibinfo{author}{\bibfnamefont{A.~H.} \bibnamefont{Guth}},
  \bibinfo{journal}{\prd} \textbf{\bibinfo{volume}{23}}, \bibinfo{pages}{347}
  (\bibinfo{year}{1981}).

\bibitem[{\citenamefont{Sato}(1981)}]{Sato:1981}
\bibinfo{author}{\bibfnamefont{K.}~\bibnamefont{Sato}},
  \bibinfo{journal}{\mnras} \textbf{\bibinfo{volume}{195}},
  \bibinfo{pages}{467} (\bibinfo{year}{1981}),
  \urlprefix\url{https://doi.org/10.1093/mnras/195.3.467}.

\bibitem[{\citenamefont{{Linde}}(1982)}]{Linde:1982}
\bibinfo{author}{\bibfnamefont{A.~D.} \bibnamefont{{Linde}}},
  \bibinfo{journal}{\prb} \textbf{\bibinfo{volume}{108}}, \bibinfo{pages}{389}
  (\bibinfo{year}{1982}).

\bibitem[{\citenamefont{Albrecht and Steinhardt}(1982)}]{Albrecht:1982}
\bibinfo{author}{\bibfnamefont{A.}~\bibnamefont{Albrecht}} \bibnamefont{and}
  \bibinfo{author}{\bibfnamefont{P.~J.} \bibnamefont{Steinhardt}},
  \bibinfo{journal}{\prl} \textbf{\bibinfo{volume}{48}}, \bibinfo{pages}{1220}
  (\bibinfo{year}{1982}),
  \urlprefix\url{https://link.aps.org/doi/10.1103/PhysRevLett.48.1220}.

\bibitem[{\citenamefont{Kamionkowski and Kovetz}(2016)}]{Kamionkowski:2015yta}
\bibinfo{author}{\bibfnamefont{M.}~\bibnamefont{Kamionkowski}}
  \bibnamefont{and} \bibinfo{author}{\bibfnamefont{E.~D.}
  \bibnamefont{Kovetz}}, \bibinfo{journal}{\araa}
  \textbf{\bibinfo{volume}{54}}, \bibinfo{pages}{227} (\bibinfo{year}{2016}),
  \eprint{1510.06042}.

\bibitem[{\citenamefont{Komatsu}(2022)}]{Komatsu:2022:review}
\bibinfo{author}{\bibfnamefont{E.}~\bibnamefont{Komatsu}},
  \bibinfo{journal}{Nature Rev. Phys.} \textbf{\bibinfo{volume}{4}},
  \bibinfo{pages}{452} (\bibinfo{year}{2022}), \eprint{2202.13919}.

\bibitem[{\citenamefont{{BICEP/Keck Collaboration: P. A. R. Ade}
  et~al.}(2021)\citenamefont{{BICEP/Keck Collaboration: P. A. R. Ade}, {Ahmed},
  {Amiri}, {Barkats}, {Basu Thakur}, {Beck}, {Bischoff}, {Bock}, {Boenish},
  {Bullock} et~al.}}]{BK13}
\bibinfo{author}{\bibnamefont{{BICEP/Keck Collaboration: P. A. R. Ade}}},
  \bibinfo{author}{\bibfnamefont{Z.}~\bibnamefont{{Ahmed}}},
  \bibinfo{author}{\bibfnamefont{M.}~\bibnamefont{{Amiri}}},
  \bibinfo{author}{\bibfnamefont{D.}~\bibnamefont{{Barkats}}},
  \bibinfo{author}{\bibfnamefont{R.}~\bibnamefont{{Basu Thakur}}},
  \bibinfo{author}{\bibfnamefont{D.}~\bibnamefont{{Beck}}},
  \bibinfo{author}{\bibfnamefont{C.}~\bibnamefont{{Bischoff}}},
  \bibinfo{author}{\bibfnamefont{J.~J.} \bibnamefont{{Bock}}},
  \bibinfo{author}{\bibfnamefont{H.}~\bibnamefont{{Boenish}}},
  \bibinfo{author}{\bibfnamefont{E.}~\bibnamefont{{Bullock}}},
  \bibnamefont{et~al.}, \bibinfo{journal}{\prl} \textbf{\bibinfo{volume}{127}},
  \bibinfo{pages}{151301} (\bibinfo{year}{2021}), \eprint{2110.00483}.

\bibitem[{\citenamefont{Tristram et~al.}(2022)}]{Tristram:2021}
\bibinfo{author}{\bibfnamefont{M.}~\bibnamefont{Tristram}}
  \bibnamefont{et~al.}, \bibinfo{journal}{\prd} \textbf{\bibinfo{volume}{105}},
  \bibinfo{pages}{083524} (\bibinfo{year}{2022}), \eprint{2112.07961}.

\bibitem[{\citenamefont{Campeti and Komatsu}(2022)}]{Campeti:2022}
\bibinfo{author}{\bibfnamefont{P.}~\bibnamefont{Campeti}} \bibnamefont{and}
  \bibinfo{author}{\bibfnamefont{E.}~\bibnamefont{Komatsu}},
  \bibinfo{journal}{\apj} \textbf{\bibinfo{volume}{941}}, \bibinfo{pages}{110}
  (\bibinfo{year}{2022}), \eprint{2205.05617}.

\bibitem[{\citenamefont{Moncelsi et~al.}(2020)\citenamefont{Moncelsi, Ade,
  Ahmed, Amiri, Barkats, {Basu Thakur}, Bischoff, Bock, Buza, Cheshire
  et~al.}}]{BICEPArray}
\bibinfo{author}{\bibfnamefont{L.}~\bibnamefont{Moncelsi}},
  \bibinfo{author}{\bibfnamefont{P.~A.~R.} \bibnamefont{Ade}},
  \bibinfo{author}{\bibfnamefont{Z.}~\bibnamefont{Ahmed}},
  \bibinfo{author}{\bibfnamefont{M.}~\bibnamefont{Amiri}},
  \bibinfo{author}{\bibfnamefont{D.}~\bibnamefont{Barkats}},
  \bibinfo{author}{\bibfnamefont{R.}~\bibnamefont{{Basu Thakur}}},
  \bibinfo{author}{\bibfnamefont{C.~A.} \bibnamefont{Bischoff}},
  \bibinfo{author}{\bibfnamefont{J.~J.} \bibnamefont{Bock}},
  \bibinfo{author}{\bibfnamefont{V.}~\bibnamefont{Buza}},
  \bibinfo{author}{\bibfnamefont{J.~R.} \bibnamefont{Cheshire}},
  \bibnamefont{et~al.}, \bibinfo{journal}{Proc. SPIE Int. Soc. Opt. Eng.}
  \textbf{\bibinfo{volume}{11453}}, \bibinfo{pages}{1145314}
  (\bibinfo{year}{2020}), \eprint{2012.04047}.

\bibitem[{\citenamefont{{Suzuki} et~al.}(2016)\citenamefont{{Suzuki}, {Ade},
  {Akiba}, {Aleman}, {Arnold}, {Baccigalupi}, {Barch}, {Barron}, {Bender},
  {Boettger} et~al.}}]{SimonsArray}
\bibinfo{author}{\bibfnamefont{A.}~\bibnamefont{{Suzuki}}},
  \bibinfo{author}{\bibfnamefont{P.}~\bibnamefont{{Ade}}},
  \bibinfo{author}{\bibfnamefont{Y.}~\bibnamefont{{Akiba}}},
  \bibinfo{author}{\bibfnamefont{C.}~\bibnamefont{{Aleman}}},
  \bibinfo{author}{\bibfnamefont{K.}~\bibnamefont{{Arnold}}},
  \bibinfo{author}{\bibfnamefont{C.}~\bibnamefont{{Baccigalupi}}},
  \bibinfo{author}{\bibfnamefont{B.}~\bibnamefont{{Barch}}},
  \bibinfo{author}{\bibfnamefont{D.}~\bibnamefont{{Barron}}},
  \bibinfo{author}{\bibfnamefont{A.}~\bibnamefont{{Bender}}},
  \bibinfo{author}{\bibfnamefont{D.}~\bibnamefont{{Boettger}}},
  \bibnamefont{et~al.}, \bibinfo{journal}{\jltp}
  \textbf{\bibinfo{volume}{184}}, \bibinfo{pages}{805} (\bibinfo{year}{2016}),
  \eprint{1512.07299}.

\bibitem[{\citenamefont{{The Simons Observatory
  Collaboration}}(2019)}]{SimonsObservatory}
\bibinfo{author}{\bibnamefont{{The Simons Observatory Collaboration}}},
  \bibinfo{journal}{\jcap} \textbf{\bibinfo{volume}{02}}, \bibinfo{pages}{056}
  (\bibinfo{year}{2019}), \eprint{1808.07445}.

\bibitem[{\citenamefont{{LiteBIRD Collaboration}}(2022)}]{LiteBIRD:2022:PTEP}
\bibinfo{author}{\bibnamefont{{LiteBIRD Collaboration}}},
  \bibinfo{journal}{\ptep} \textbf{\bibinfo{volume}{2023}},
  \bibinfo{pages}{042F01} (\bibinfo{year}{2022}), \eprint{2202.02773}.

\bibitem[{\citenamefont{{CMB-S4 Collaboration: K. Abazajian}
  et~al.}(2019)\citenamefont{{CMB-S4 Collaboration: K. Abazajian}, {Addison},
  {Adshead}, {Ahmed}, {Allen}, {Alonso}, {Alvarez}, {Anderson}, {Arnold},
  {Baccigalupi} et~al.}}]{CMBS4}
\bibinfo{author}{\bibnamefont{{CMB-S4 Collaboration: K. Abazajian}}},
  \bibinfo{author}{\bibfnamefont{G.}~\bibnamefont{{Addison}}},
  \bibinfo{author}{\bibfnamefont{P.}~\bibnamefont{{Adshead}}},
  \bibinfo{author}{\bibfnamefont{Z.}~\bibnamefont{{Ahmed}}},
  \bibinfo{author}{\bibfnamefont{S.~W.} \bibnamefont{{Allen}}},
  \bibinfo{author}{\bibfnamefont{D.}~\bibnamefont{{Alonso}}},
  \bibinfo{author}{\bibfnamefont{M.}~\bibnamefont{{Alvarez}}},
  \bibinfo{author}{\bibfnamefont{A.}~\bibnamefont{{Anderson}}},
  \bibinfo{author}{\bibfnamefont{K.~S.} \bibnamefont{{Arnold}}},
  \bibinfo{author}{\bibfnamefont{C.}~\bibnamefont{{Baccigalupi}}},
  \bibnamefont{et~al.} (\bibinfo{year}{2019}), \eprint{1907.04473}.

\bibitem[{\citenamefont{{CMB-S4 Collaboration: K. Abazajian}
  et~al.}(2022)\citenamefont{{CMB-S4 Collaboration: K. Abazajian}, {Addison},
  {Adshead}, {Ahmed}, {Akerib}, {Ali}, {Allen}, {Alonso}, {Alvarez}, {Amin}
  et~al.}}]{CMBS4:r-forecast}
\bibinfo{author}{\bibnamefont{{CMB-S4 Collaboration: K. Abazajian}}},
  \bibinfo{author}{\bibfnamefont{G.~E.} \bibnamefont{{Addison}}},
  \bibinfo{author}{\bibfnamefont{P.}~\bibnamefont{{Adshead}}},
  \bibinfo{author}{\bibfnamefont{Z.}~\bibnamefont{{Ahmed}}},
  \bibinfo{author}{\bibfnamefont{D.}~\bibnamefont{{Akerib}}},
  \bibinfo{author}{\bibfnamefont{A.}~\bibnamefont{{Ali}}},
  \bibinfo{author}{\bibfnamefont{S.~W.} \bibnamefont{{Allen}}},
  \bibinfo{author}{\bibfnamefont{D.}~\bibnamefont{{Alonso}}},
  \bibinfo{author}{\bibfnamefont{M.}~\bibnamefont{{Alvarez}}},
  \bibinfo{author}{\bibfnamefont{M.~A.} \bibnamefont{{Amin}}},
  \bibnamefont{et~al.}, \bibinfo{journal}{\apj} \textbf{\bibinfo{volume}{926}},
  \bibinfo{pages}{54} (\bibinfo{year}{2022}), \eprint{2008.12619}.

\bibitem[{\citenamefont{{BICEP2 Collaboration}}(2014)}]{B2I}
\bibinfo{author}{\bibnamefont{{BICEP2 Collaboration}}}, \bibinfo{journal}{\prl}
  \textbf{\bibinfo{volume}{112}}, \bibinfo{pages}{241101}
  (\bibinfo{year}{2014}).

\bibitem[{\citenamefont{{BICEP/Keck Collaboration: P.~A.~R. Ade}
  et~al.}(2023)\citenamefont{{BICEP/Keck Collaboration: P.~A.~R. Ade}, {Ahmed},
  {Amiri}, {Barkats}, {Thakur}, {Bischoff}, {Beck}, {Bock}, {Boenish},
  {Bullock} et~al.}}]{BK-XVI:FG}
\bibinfo{author}{\bibnamefont{{BICEP/Keck Collaboration: P.~A.~R. Ade}}},
  \bibinfo{author}{\bibfnamefont{Z.}~\bibnamefont{{Ahmed}}},
  \bibinfo{author}{\bibfnamefont{M.}~\bibnamefont{{Amiri}}},
  \bibinfo{author}{\bibfnamefont{D.}~\bibnamefont{{Barkats}}},
  \bibinfo{author}{\bibfnamefont{R.~B.} \bibnamefont{{Thakur}}},
  \bibinfo{author}{\bibfnamefont{C.~A.} \bibnamefont{{Bischoff}}},
  \bibinfo{author}{\bibfnamefont{D.}~\bibnamefont{{Beck}}},
  \bibinfo{author}{\bibfnamefont{J.~J.} \bibnamefont{{Bock}}},
  \bibinfo{author}{\bibfnamefont{H.}~\bibnamefont{{Boenish}}},
  \bibinfo{author}{\bibfnamefont{E.}~\bibnamefont{{Bullock}}},
  \bibnamefont{et~al.}, \bibinfo{journal}{\apj} \textbf{\bibinfo{volume}{945}},
  \bibinfo{pages}{72} (\bibinfo{year}{2023}), \eprint{2210.05684}.

\bibitem[{\citenamefont{Eriksen et~al.}(2008)\citenamefont{Eriksen, Jewell,
  Dickinson, Banday, Gorski, and Lawrence}}]{Eriksen:2007}
\bibinfo{author}{\bibfnamefont{H.~K.} \bibnamefont{Eriksen}},
  \bibinfo{author}{\bibfnamefont{J.~B.} \bibnamefont{Jewell}},
  \bibinfo{author}{\bibfnamefont{C.}~\bibnamefont{Dickinson}},
  \bibinfo{author}{\bibfnamefont{A.~J.} \bibnamefont{Banday}},
  \bibinfo{author}{\bibfnamefont{K.~M.} \bibnamefont{Gorski}},
  \bibnamefont{and} \bibinfo{author}{\bibfnamefont{C.~R.}
  \bibnamefont{Lawrence}}, \bibinfo{journal}{\apj}
  \textbf{\bibinfo{volume}{676}}, \bibinfo{pages}{10} (\bibinfo{year}{2008}),
  \eprint{0709.1058}.

\bibitem[{\citenamefont{{Dunkley} et~al.}(2009)\citenamefont{{Dunkley},
  {Amblard}, {Baccigalupi}, {Betoule}, {Chuss}, {Cooray}, {Delabrouille},
  {Dickinson}, {Dobler}, {Dotson} et~al.}}]{Dunkley:2008}
\bibinfo{author}{\bibfnamefont{J.}~\bibnamefont{{Dunkley}}},
  \bibinfo{author}{\bibfnamefont{A.}~\bibnamefont{{Amblard}}},
  \bibinfo{author}{\bibfnamefont{C.}~\bibnamefont{{Baccigalupi}}},
  \bibinfo{author}{\bibfnamefont{M.}~\bibnamefont{{Betoule}}},
  \bibinfo{author}{\bibfnamefont{D.}~\bibnamefont{{Chuss}}},
  \bibinfo{author}{\bibfnamefont{A.}~\bibnamefont{{Cooray}}},
  \bibinfo{author}{\bibfnamefont{J.}~\bibnamefont{{Delabrouille}}},
  \bibinfo{author}{\bibfnamefont{C.}~\bibnamefont{{Dickinson}}},
  \bibinfo{author}{\bibfnamefont{G.}~\bibnamefont{{Dobler}}},
  \bibinfo{author}{\bibfnamefont{J.}~\bibnamefont{{Dotson}}},
  \bibnamefont{et~al.}, \textbf{\bibinfo{volume}{1141}}, \bibinfo{pages}{222}
  (\bibinfo{year}{2009}), \eprint{0811.3915}.

\bibitem[{\citenamefont{Betoule et~al.}(2009)\citenamefont{Betoule, Pierpaoli,
  Delabrouille, Le~Jeune, and Cardoso}}]{Betoule:2009}
\bibinfo{author}{\bibfnamefont{M.}~\bibnamefont{Betoule}},
  \bibinfo{author}{\bibfnamefont{E.}~\bibnamefont{Pierpaoli}},
  \bibinfo{author}{\bibfnamefont{J.}~\bibnamefont{Delabrouille}},
  \bibinfo{author}{\bibfnamefont{M.}~\bibnamefont{Le~Jeune}}, \bibnamefont{and}
  \bibinfo{author}{\bibfnamefont{J.-F.} \bibnamefont{Cardoso}},
  \bibinfo{journal}{Astronomy and Astrophysics} \textbf{\bibinfo{volume}{503}},
  \bibinfo{pages}{691} (\bibinfo{year}{2009}), \eprint{0901.1056}.

\bibitem[{\citenamefont{Katayama and Komatsu}(2011)}]{Katayama:2011eh}
\bibinfo{author}{\bibfnamefont{N.}~\bibnamefont{Katayama}} \bibnamefont{and}
  \bibinfo{author}{\bibfnamefont{E.}~\bibnamefont{Komatsu}},
  \bibinfo{journal}{\apj} \textbf{\bibinfo{volume}{737}}, \bibinfo{pages}{78}
  (\bibinfo{year}{2011}), \eprint{1101.5210}.

\bibitem[{\citenamefont{Ichiki}(2014)}]{Ichiki:2014}
\bibinfo{author}{\bibfnamefont{K.}~\bibnamefont{Ichiki}},
  \bibinfo{journal}{\ptep} \textbf{\bibinfo{volume}{06}}, \bibinfo{pages}{B109}
  (\bibinfo{year}{2014}).

\bibitem[{\citenamefont{Chluba et~al.}(2017)\citenamefont{Chluba, Hill, and
  Abitbol}}]{Chluba:2017}
\bibinfo{author}{\bibfnamefont{J.}~\bibnamefont{Chluba}},
  \bibinfo{author}{\bibfnamefont{J.~C.} \bibnamefont{Hill}}, \bibnamefont{and}
  \bibinfo{author}{\bibfnamefont{M.~H.} \bibnamefont{Abitbol}},
  \bibinfo{journal}{\mnras} \textbf{\bibinfo{volume}{472}},
  \bibinfo{pages}{1195} (\bibinfo{year}{2017}), \eprint{1701.00274}.

\bibitem[{\citenamefont{Azzoni et~al.}(2021)\citenamefont{Azzoni, Abitbol,
  Alonso, Gough, Katayama, and Matsumura}}]{Azzoni:2020}
\bibinfo{author}{\bibfnamefont{S.}~\bibnamefont{Azzoni}},
  \bibinfo{author}{\bibfnamefont{M.~H.} \bibnamefont{Abitbol}},
  \bibinfo{author}{\bibfnamefont{D.}~\bibnamefont{Alonso}},
  \bibinfo{author}{\bibfnamefont{A.}~\bibnamefont{Gough}},
  \bibinfo{author}{\bibfnamefont{N.}~\bibnamefont{Katayama}}, \bibnamefont{and}
  \bibinfo{author}{\bibfnamefont{T.}~\bibnamefont{Matsumura}},
  \bibinfo{journal}{\jcap} \textbf{\bibinfo{volume}{05}}, \bibinfo{pages}{047}
  (\bibinfo{year}{2021}), \eprint{2011.11575}.

\bibitem[{\citenamefont{de~la Hoz et~al.}(2020)\citenamefont{de~la Hoz, Vielva,
  Barreiro, and Mart\'\i{}nez-Gonz\'alez}}]{delaHoz:2020}
\bibinfo{author}{\bibfnamefont{E.}~\bibnamefont{de~la Hoz}},
  \bibinfo{author}{\bibfnamefont{P.}~\bibnamefont{Vielva}},
  \bibinfo{author}{\bibfnamefont{R.~B.} \bibnamefont{Barreiro}},
  \bibnamefont{and}
  \bibinfo{author}{\bibfnamefont{E.}~\bibnamefont{Mart\'\i{}nez-Gonz\'alez}},
  \bibinfo{journal}{\jcap} \textbf{\bibinfo{volume}{06}}, \bibinfo{pages}{006}
  (\bibinfo{year}{2020}), \eprint{2002.12206}.

\bibitem[{\citenamefont{Remazeilles et~al.}(2021)\citenamefont{Remazeilles,
  Rotti, and Chluba}}]{Remazeilles:2020}
\bibinfo{author}{\bibfnamefont{M.}~\bibnamefont{Remazeilles}},
  \bibinfo{author}{\bibfnamefont{A.}~\bibnamefont{Rotti}}, \bibnamefont{and}
  \bibinfo{author}{\bibfnamefont{J.}~\bibnamefont{Chluba}},
  \bibinfo{journal}{\mnras} \textbf{\bibinfo{volume}{503}},
  \bibinfo{pages}{2478} (\bibinfo{year}{2021}), \eprint{2006.08628}.

\bibitem[{\citenamefont{Puglisi et~al.}(2022)\citenamefont{Puglisi, Mihaylov,
  Panopoulou, Poletti, Errard, Puglisi, and Vianello}}]{Puglisi:2021}
\bibinfo{author}{\bibfnamefont{G.}~\bibnamefont{Puglisi}},
  \bibinfo{author}{\bibfnamefont{G.}~\bibnamefont{Mihaylov}},
  \bibinfo{author}{\bibfnamefont{G.~V.} \bibnamefont{Panopoulou}},
  \bibinfo{author}{\bibfnamefont{D.}~\bibnamefont{Poletti}},
  \bibinfo{author}{\bibfnamefont{J.}~\bibnamefont{Errard}},
  \bibinfo{author}{\bibfnamefont{P.~A.} \bibnamefont{Puglisi}},
  \bibnamefont{and} \bibinfo{author}{\bibfnamefont{G.}~\bibnamefont{Vianello}},
  \bibinfo{journal}{\mnras} \textbf{\bibinfo{volume}{511}},
  \bibinfo{pages}{2052} (\bibinfo{year}{2022}), \eprint{2109.11562}.

\bibitem[{\citenamefont{Delouis et~al.}(2022)\citenamefont{Delouis, Allys,
  Gauvrit, and Boulanger}}]{Delouis:2022}
\bibinfo{author}{\bibfnamefont{J.~M.} \bibnamefont{Delouis}},
  \bibinfo{author}{\bibfnamefont{E.}~\bibnamefont{Allys}},
  \bibinfo{author}{\bibfnamefont{E.}~\bibnamefont{Gauvrit}}, \bibnamefont{and}
  \bibinfo{author}{\bibfnamefont{F.}~\bibnamefont{Boulanger}},
  \bibinfo{journal}{\aap} \textbf{\bibinfo{volume}{668}}, \bibinfo{pages}{A122}
  (\bibinfo{year}{2022}), \eprint{2207.12527}.

\bibitem[{\citenamefont{Minami and Ichiki}(2022)}]{Minami&Ichiki:2022}
\bibinfo{author}{\bibfnamefont{Y.}~\bibnamefont{Minami}} \bibnamefont{and}
  \bibinfo{author}{\bibfnamefont{K.}~\bibnamefont{Ichiki}},
  \bibinfo{journal}{\ptep} \textbf{\bibinfo{volume}{2023}},
  \bibinfo{pages}{033E01} (\bibinfo{year}{2022}), \eprint{2212.01773}.

\bibitem[{\citenamefont{Rahman et~al.}(2022)\citenamefont{Rahman, Chingangbam,
  and Ghosh}}]{Rahman:2022}
\bibinfo{author}{\bibfnamefont{F.}~\bibnamefont{Rahman}},
  \bibinfo{author}{\bibfnamefont{P.}~\bibnamefont{Chingangbam}},
  \bibnamefont{and} \bibinfo{author}{\bibfnamefont{T.}~\bibnamefont{Ghosh}}
  (\bibinfo{year}{2022}), \eprint{2212.06076}.

\bibitem[{\citenamefont{Vacher et~al.}(2023)\citenamefont{Vacher, Chluba,
  Aumont, Rotti, and Montier}}]{Vacher:2022}
\bibinfo{author}{\bibfnamefont{L.}~\bibnamefont{Vacher}},
  \bibinfo{author}{\bibfnamefont{J.}~\bibnamefont{Chluba}},
  \bibinfo{author}{\bibfnamefont{J.}~\bibnamefont{Aumont}},
  \bibinfo{author}{\bibfnamefont{A.}~\bibnamefont{Rotti}}, \bibnamefont{and}
  \bibinfo{author}{\bibfnamefont{L.}~\bibnamefont{Montier}},
  \bibinfo{journal}{\aap} \textbf{\bibinfo{volume}{669}}, \bibinfo{pages}{A5}
  (\bibinfo{year}{2023}), \eprint{2205.01049}.

\bibitem[{\citenamefont{Carones et~al.}(2022)\citenamefont{Carones, Migliaccio,
  Puglisi, Baccigalupi, Marinucci, Vittorio, and Poletti}}]{Carones:2022}
\bibinfo{author}{\bibfnamefont{A.}~\bibnamefont{Carones}},
  \bibinfo{author}{\bibfnamefont{M.}~\bibnamefont{Migliaccio}},
  \bibinfo{author}{\bibfnamefont{G.}~\bibnamefont{Puglisi}},
  \bibinfo{author}{\bibfnamefont{C.}~\bibnamefont{Baccigalupi}},
  \bibinfo{author}{\bibfnamefont{D.}~\bibnamefont{Marinucci}},
  \bibinfo{author}{\bibfnamefont{N.}~\bibnamefont{Vittorio}}, \bibnamefont{and}
  \bibinfo{author}{\bibfnamefont{D.}~\bibnamefont{Poletti}}
  (\bibinfo{year}{2022}), \eprint{2212.04456}.

\bibitem[{\citenamefont{Hu and Okamoto}(2002)}]{HuOkamoto:2001}
\bibinfo{author}{\bibfnamefont{W.}~\bibnamefont{Hu}} \bibnamefont{and}
  \bibinfo{author}{\bibfnamefont{T.}~\bibnamefont{Okamoto}},
  \bibinfo{journal}{\apj} \textbf{\bibinfo{volume}{574}}, \bibinfo{pages}{566}
  (\bibinfo{year}{2002}), \eprint{astro-ph/0111606}.

\bibitem[{\citenamefont{Zaldarriaga and Seljak}(1997)}]{Zaldarriaga:1996:EBdef}
\bibinfo{author}{\bibfnamefont{M.}~\bibnamefont{Zaldarriaga}} \bibnamefont{and}
  \bibinfo{author}{\bibfnamefont{U.}~\bibnamefont{Seljak}},
  \bibinfo{journal}{\prd} \textbf{\bibinfo{volume}{D55}}, \bibinfo{pages}{1830}
  (\bibinfo{year}{1997}), \eprint{astro-ph/9609170}.

\bibitem[{\citenamefont{Lewis and Challinor}(2006)}]{Lewis:2006fu}
\bibinfo{author}{\bibfnamefont{A.}~\bibnamefont{Lewis}} \bibnamefont{and}
  \bibinfo{author}{\bibfnamefont{A.}~\bibnamefont{Challinor}},
  \bibinfo{journal}{Phys. Rep.} \textbf{\bibinfo{volume}{429}},
  \bibinfo{pages}{1} (\bibinfo{year}{2006}), \eprint{astro-ph/0601594}.

\bibitem[{\citenamefont{Zaldarriaga and Seljak}(1998)}]{Zaldarriaga:1998:LensB}
\bibinfo{author}{\bibfnamefont{M.}~\bibnamefont{Zaldarriaga}} \bibnamefont{and}
  \bibinfo{author}{\bibfnamefont{U.}~\bibnamefont{Seljak}},
  \bibinfo{journal}{\prd} \textbf{\bibinfo{volume}{58}},
  \bibinfo{pages}{023003} (\bibinfo{year}{1998}), \eprint{astro-ph/9803150}.

\bibitem[{\citenamefont{Fabbian et~al.}(2019)\citenamefont{Fabbian, Lewis, and
  Beck}}]{Fabbian:2019}
\bibinfo{author}{\bibfnamefont{G.}~\bibnamefont{Fabbian}},
  \bibinfo{author}{\bibfnamefont{A.}~\bibnamefont{Lewis}}, \bibnamefont{and}
  \bibinfo{author}{\bibfnamefont{D.}~\bibnamefont{Beck}},
  \bibinfo{journal}{\jcap} \textbf{\bibinfo{volume}{10}}, \bibinfo{pages}{057}
  (\bibinfo{year}{2019}), \eprint{1906.08760}.

\bibitem[{\citenamefont{Okamoto and Hu}(2003)}]{OkamotoHu:quad}
\bibinfo{author}{\bibfnamefont{T.}~\bibnamefont{Okamoto}} \bibnamefont{and}
  \bibinfo{author}{\bibfnamefont{W.}~\bibnamefont{Hu}}, \bibinfo{journal}{\prd}
  \textbf{\bibinfo{volume}{67}}, \bibinfo{pages}{083002}
  (\bibinfo{year}{2003}), \eprint{astro-ph/0301031}.

\bibitem[{\citenamefont{Seljak and Hirata}(2004)}]{Seljak:2003pn}
\bibinfo{author}{\bibfnamefont{U.}~\bibnamefont{Seljak}} \bibnamefont{and}
  \bibinfo{author}{\bibfnamefont{C.~M.} \bibnamefont{Hirata}},
  \bibinfo{journal}{\prd} \textbf{\bibinfo{volume}{69}},
  \bibinfo{pages}{043005} (\bibinfo{year}{2004}), \eprint{astro-ph/0310163}.

\bibitem[{\citenamefont{Baleato~Lizancos
  et~al.}(2021)\citenamefont{Baleato~Lizancos, Challinor, and
  Carron}}]{BaleatoLizancos:2020:delens-bias}
\bibinfo{author}{\bibfnamefont{A.}~\bibnamefont{Baleato~Lizancos}},
  \bibinfo{author}{\bibfnamefont{A.}~\bibnamefont{Challinor}},
  \bibnamefont{and} \bibinfo{author}{\bibfnamefont{J.}~\bibnamefont{Carron}},
  \bibinfo{journal}{\jcap} \textbf{\bibinfo{volume}{03}}, \bibinfo{pages}{016}
  (\bibinfo{year}{2021}), \eprint{2007.01622}.

\bibitem[{\citenamefont{Sherwin and Schmittfull}(2015)}]{Sherwin:2015}
\bibinfo{author}{\bibfnamefont{B.~D.} \bibnamefont{Sherwin}} \bibnamefont{and}
  \bibinfo{author}{\bibfnamefont{M.}~\bibnamefont{Schmittfull}},
  \bibinfo{journal}{\prd} \textbf{\bibinfo{volume}{92}},
  \bibinfo{pages}{043005} (\bibinfo{year}{2015}), \eprint{1502.05356}.

\bibitem[{\citenamefont{Errard et~al.}(2016)\citenamefont{Errard, Feeney,
  Peiris, and Jaffe}}]{Errard:2016:FG}
\bibinfo{author}{\bibfnamefont{J.}~\bibnamefont{Errard}},
  \bibinfo{author}{\bibfnamefont{S.~M.} \bibnamefont{Feeney}},
  \bibinfo{author}{\bibfnamefont{H.~V.} \bibnamefont{Peiris}},
  \bibnamefont{and} \bibinfo{author}{\bibfnamefont{A.~H.} \bibnamefont{Jaffe}},
  \bibinfo{journal}{\jcap} \textbf{\bibinfo{volume}{03}}, \bibinfo{pages}{052}
  (\bibinfo{year}{2016}).

\bibitem[{\citenamefont{Teng et~al.}(2011)\citenamefont{Teng, Kuo, and
  Wu}}]{Teng:2011xc}
\bibinfo{author}{\bibfnamefont{W.-H.} \bibnamefont{Teng}},
  \bibinfo{author}{\bibfnamefont{C.-L.} \bibnamefont{Kuo}}, \bibnamefont{and}
  \bibinfo{author}{\bibfnamefont{J.-H.~P.} \bibnamefont{Wu}}
  (\bibinfo{year}{2011}), \eprint{1102.5729}.

\bibitem[{\citenamefont{Namikawa and Nagata}(2014)}]{Namikawa:2014:patchwork}
\bibinfo{author}{\bibfnamefont{T.}~\bibnamefont{Namikawa}} \bibnamefont{and}
  \bibinfo{author}{\bibfnamefont{R.}~\bibnamefont{Nagata}},
  \bibinfo{journal}{\jcap} \textbf{\bibinfo{volume}{09}}, \bibinfo{pages}{009}
  (\bibinfo{year}{2014}), \eprint{1405.6568}.

\bibitem[{\citenamefont{Smith et~al.}(2012)\citenamefont{Smith, Hanson,
  LoVerde, Hirata, and Zahn}}]{Smith:2010gu}
\bibinfo{author}{\bibfnamefont{K.~M.} \bibnamefont{Smith}},
  \bibinfo{author}{\bibfnamefont{D.}~\bibnamefont{Hanson}},
  \bibinfo{author}{\bibfnamefont{M.}~\bibnamefont{LoVerde}},
  \bibinfo{author}{\bibfnamefont{C.~M.} \bibnamefont{Hirata}},
  \bibnamefont{and} \bibinfo{author}{\bibfnamefont{O.}~\bibnamefont{Zahn}},
  \bibinfo{journal}{\jcap} \textbf{\bibinfo{volume}{06}}, \bibinfo{pages}{014}
  (\bibinfo{year}{2012}), \eprint{1010.0048}.

\bibitem[{\citenamefont{Namikawa et~al.}(2022)\citenamefont{Namikawa, {Baleato
  Lizancos}, Robertson, Sherwin, Challinor et~al.}}]{Namikawa:2021:so-delens}
\bibinfo{author}{\bibfnamefont{T.}~\bibnamefont{Namikawa}},
  \bibinfo{author}{\bibfnamefont{A.}~\bibnamefont{{Baleato Lizancos}}},
  \bibinfo{author}{\bibfnamefont{N.}~\bibnamefont{Robertson}},
  \bibinfo{author}{\bibfnamefont{B.~D.} \bibnamefont{Sherwin}},
  \bibinfo{author}{\bibfnamefont{A.}~\bibnamefont{Challinor}},
  \bibnamefont{et~al.}, \bibinfo{journal}{\prd} \textbf{\bibinfo{volume}{105}},
  \bibinfo{pages}{023511} (\bibinfo{year}{2022}), \eprint{2110.09730}.

\bibitem[{\citenamefont{{LSST Dark Energy Science Collaboration}}(2012)}]{LSST}
\bibinfo{author}{\bibnamefont{{LSST Dark Energy Science Collaboration}}}
  (\bibinfo{year}{2012}), \eprint{arXiv:1211.0310}.

\bibitem[{\citenamefont{Aurlien et~al.}(2022)}]{Aurlien:2022:FG}
\bibinfo{author}{\bibfnamefont{R.}~\bibnamefont{Aurlien}} \bibnamefont{et~al.},
  \bibinfo{journal}{\jcap} \textbf{\bibinfo{volume}{06}}, \bibinfo{pages}{034}
  (\bibinfo{year}{2022}), \eprint{2211.14342}.

\bibitem[{\citenamefont{Baleato~Lizancos and
  Ferraro}(2022)}]{BaleatoLizancos:2022:extFGs}
\bibinfo{author}{\bibfnamefont{A.}~\bibnamefont{Baleato~Lizancos}}
  \bibnamefont{and} \bibinfo{author}{\bibfnamefont{S.}~\bibnamefont{Ferraro}},
  \bibinfo{journal}{\prd} \textbf{\bibinfo{volume}{106}},
  \bibinfo{pages}{063534} (\bibinfo{year}{2022}), \eprint{2205.09000}.

\bibitem[{\citenamefont{Namikawa and Takahashi}(2014)}]{Namikawa:2013:bhepol}
\bibinfo{author}{\bibfnamefont{T.}~\bibnamefont{Namikawa}} \bibnamefont{and}
  \bibinfo{author}{\bibfnamefont{R.}~\bibnamefont{Takahashi}},
  \bibinfo{journal}{\mnras} \textbf{\bibinfo{volume}{438}},
  \bibinfo{pages}{1507} (\bibinfo{year}{2014}), \eprint{1310.2372}.

\bibitem[{\citenamefont{Sailer et~al.}(2023)\citenamefont{Sailer, Ferraro, and
  Schaan}}]{Sailer:2022}
\bibinfo{author}{\bibfnamefont{N.}~\bibnamefont{Sailer}},
  \bibinfo{author}{\bibfnamefont{S.}~\bibnamefont{Ferraro}}, \bibnamefont{and}
  \bibinfo{author}{\bibfnamefont{E.}~\bibnamefont{Schaan}},
  \bibinfo{journal}{\prd} \textbf{\bibinfo{volume}{107}},
  \bibinfo{pages}{023504} (\bibinfo{year}{2023}), \eprint{2211.03786}.

\bibitem[{\citenamefont{Namikawa and Nagata}(2015)}]{Namikawa:2015:delens}
\bibinfo{author}{\bibfnamefont{T.}~\bibnamefont{Namikawa}} \bibnamefont{and}
  \bibinfo{author}{\bibfnamefont{R.}~\bibnamefont{Nagata}},
  \bibinfo{journal}{\jcap} \textbf{\bibinfo{volume}{10}}, \bibinfo{pages}{004}
  (\bibinfo{year}{2015}), \eprint{1506.09209}.

\end{thebibliography}

\end{document}